\documentclass[useAMS,usenatbib]{mn2e}

\usepackage{graphicx,amssymb}

\newcommand{\fsize}{0.45\textwidth}
\newcommand{\cs}{$\chi^2$}

\title[TrES-1b Transit Analysis]{An analysis of the transit times of TrES-1b}

\author[Steffen and Agol]{Jason H. Steffen\thanks{steffenATastro.washington.edu} and Eric Agol\\ Physics Department, University of Washington, Box 351560, Seattle, WA 98195}

\begin{document}

\maketitle

\begin{abstract}
The presence of a second planet in a known, transiting-planet system will cause the time between transits to vary.  These variations can be used to constrain the orbital elements and mass of the perturbing planet.  We analyse the set of transit times of the TrES-1 system given in \cite{char05}.  We find no convincing evidence for a second planet in the TrES-1 system from that data.  By further analysis, we constrain the mass that a perturbing planet could have as a function of the semi-major axis ratio of the two planets and the eccentricity of the perturbing planet.  Near low-order, mean-motion resonances (within $\sim 1\%$ fractional deviation), we find that a secondary planet must generally have a mass comparable to or less than the mass of the Earth--showing that this data is the first to have sensitivity to sub Earth-mass planets.  We compare the sensitivity of this technique to the mass of the perturbing planet with future, high-precision radial velocity measurements.
\end{abstract}



\section{Introduction}

Since the first discovery, a decade ago, of a planet orbiting a distant main-sequence star~\citep{mayor95}, several search techniques have been employed to identify additional extrasolar planets.  These techniques include radial velocity~\citep{marcy05,moutou05}, astrometry \citep{ford03,glind03,pravdo04,gouda05,lattanzi05}, planetary microlensing~\citep{albrow00,tsapras03}, and planetary transits~\citep{konacki03,bouchy04,pont04,char05,konacki05}.  A recent addition to the repertoire of planet search techniques consists of looking for additional planets in a known, transiting system by analyzing the variation in the time between planetary transits.  These transit timing variations (TTV) can be used to constrain the orbital elements of an unseen, perturbing planet, even if its mass is comparable to the mass of the Earth~\citep{escude02,agol05,holm05}.  Thus, for the near term, TTV can detect planets around main-sequence stars that are too small to detect by any other means.  This sensitivity is particularly manifest near mean-motion resonances, which recent works by \cite{thommes}, \cite{papal}, and \cite{zhou} suggest might be very common.

The study of extrasolar planetary systems allows us to address fundamental questions about mechanisms of planet formation, the prevalence of small, rocky planets like our Earth, and the evolution of multiple planetary systems.  Multiple planet systems are an important subset of extrasolar planets because they give insight into our own solar system.  TTV should prove to be an important tool in sampling the population of such planets since it has the ability to discern the presence of bodies with very little mass ($\lesssim M_{\oplus}$).

The application of TTV depends upon the discovery and monitoring of transiting planetary systems.  The first successful detection of a planetary transit~\citep{charbon00,henry00} was for a planet identified spectroscopically, HD209458b~\citep{mazeh00}.  The first extrasolar planet to be discovered from transit data was the OGLE-TR-56b system reported by \cite{konacki03}.  Existing and future searches for planetary transits, such as the COROT~\citep{bourde}, XO \citep{mccull05}, and Kepler~\citep{borucki} missions, are expected to provide many candidate systems where the timing of the transits can be analysed.

The TrES-1 planetary system, reported by \cite{alonso}, was also discovered via planetary transits.  In a recent paper by \cite{char05} (hereafter C05), the detection of thermal emission from the surface of TrES-1b was announced.  Table 1 of C05 gives the timing of 12 planetary transits.  We analyse the deviations from periodicity in those data in order to identify and constrain potential companion planets in that system.

\section{The Data and Software}

The timing data reported in C05 were derived from the 11 transits reported by \cite{alonso} with an additional transit that was observed at the IAC 80cm telescope after \cite{alonso} went to press.  One transit was excluded from their analysis because it constituted a 6-$\sigma$ departure from a constant period and because of anomalous features in the ingress and egress.  That point, if it is valid, is the most interesting point for our purposes because the TTV signal is defined by such deviations.  Consequently, we analyse two different sets of data from C05; the ``12-point'' set which includes this point, and the ``11-point'' set which does not.  This study is the first analysis of TTV as presented by \cite{agol05} and \cite{holm05}.  And, it may be the first search for planets around main-sequence stars that can probe masses smaller than the mass of the Earth.

To calculate the time of transit, we integrate the equations of motion using the Burlisch-Stoer integration algorithm.  When the transiting planet crosses the line of sight, the transit time is identified and tabulated.  These simulated transit times are compared to the observed transit times and the quality of the fit of the orbital elements of the simulated system with those of the actual system is characterized by the \cs\ statistic.  For this work we restrict our analysis to planets whose orbits are coplanar and edge-on.

To locate the set of orbital elements that gives the minimum \cs, we sequentially employ a random and then a direct minimization algorithm.  For the random portion we generate many sets of orbital elements and record the \cs\ for each.  All parameters are drawn from a uniform distribution, except for the eccentricity and mass which are logarithmically uniform; between $10^{-3.4}$ and $1.0$ for the eccentricity and between $10^{-6}$ and $10^{-2.3}$ for the ratio of the planet to stellar mass.  We chose the logarithmic distribution so as to not assume a mass scale \textit{a priori}--our limits corresponding approximately to masses too small to be seen and too large to have not been seen.  This approach agrees with the empirical distribution of masses determined by \cite{tabac02}.  The logarithmic distribution was chosen for eccentricity because tidal circularization and orbital stability constraints favor small eccentricities ($\lesssim 0.1$).  For completeness, we also conducted a study where the eccentricity of the perturbing planet began at zero.

We tabulate the two systems with the lowest \cs\ as starting points for the direct minimization algorithm.  Experiments with simulated data showed no significant improvement in the final \cs\ when more than two solutions were retained.  We assume, following the random minimization, that at least one set of orbital elements is near the global minimum of the \cs.  That is, the nearest local minimum is the global minimum and that the local gradient points toward that minimum.  Tests of this approach with simulated data confirm this assumption.  From these points we use a direct minimization algorithm to locate the bottom of the global minimum.

\section{Results}

\subsection{Search for Secondary Planets\label{search}}

We conducted a variety of searches for the best-fitting perturbing planet in the TrES-1 system.  These searches included different combinations of orbital elements for TrES-1b.  We found that any reduction in the overall \cs\ obtained by including the parameters $e$ and $\varpi$ for TrES-1b was offset by the loss of a degree of freedom.  Therefore, we report results from the search where the eccentricity of TrES-1b was fixed at zero.

We stepped through the semi-major axis ratio of the putative secondary planet and TrES-1b.  At each point we minimized over six parameters: the eccentricity, longitude of pericentre, time of pericentre passage, and mass of the secondary planet and the period and the initial longitude of TrES-1b.  The inclination and ascending node of the perturbing planet were identical to the values for TrES-1b.  We analysed the data for both interior and exterior perturbers.

\begin{figure}
\includegraphics[width=\fsize]{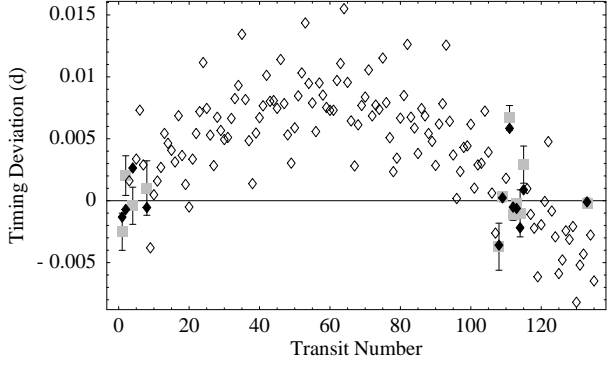}
\caption{A comparison of the TrES-1b transit timing residuals to those from the 12-point analysis for an interior perturber.  The diamonds are the timing residual at each transit, the filled diamonds denote transits where there is data, the squares with error bars are the data.  The 12-point solution has a reduced \cs\ of 2.8 on 6 degrees of freedom with ephemeris parameters $m = 0.111M_{\rm J}$, $P = 1.76d$, $e = 0.153$, $\varpi = 154.8^{\circ}$, $\tau = 2452847.4392d$.}
\label{timings1}
\end{figure}

Our analysis did not produce any promising solution; though we present one interesting case for an interior perturber found from the 12-point analysis.  This solution is not near a low-order mean-motion resonance.  Indeed, the \cs\ was generally much higher near the $j$:$j+1$ and the $j$:1 resonances than in the regions between them.  Figure \ref{timings1} compares the timing residuals for this solution with the data.  The reduced \cs\ for this system is 2.8 on $N-6$ degrees of freedom (where $N=12$ is the number of data) compared with 6.3 for no perturber ($N-2$).

This solution, while it would have been detected from RV measurements, is interesting because the average size of the timing deviations is larger than that of the data--making the variations easier to detect.  However, we suspect that while this solution is numerically valid, it is merely an artifact of the gap between the two primary epochs of observation.  For this solution, and others like it, the simulated timing residuals consist of small, short-term variations superposed on a large-amplitude, long-period variation with a period that is a multiple of the difference between the two epochs.  Several candiate systems for both the 11-point and the 12-point analyses showed similar behaviour.  Additional data, taken at a time that is not commensurate with this period would remove false solutions of this type.

From the above analysis, we conclude that there is not sufficient information in the data to uniquely and satisfactorally determine the characteristics of a secondary planet in the TrES-1 system.  This is in part because the number of model parameters is not much larger than the number of data and because the typical timing precision, $\sim 100s$, is not a sufficiently small fraction of the orbital period of the transiting planet (about $4\times 10^{-4}$) to distinguish between different solutions.  In addition, the gap in coverage appears to affect the minimization dramatically.  We believe that it is primarily responsible for the observed fact that two nearby sets of orbital elements will have very different values of \cs --a slight change in the long-term variation will cause the simulated transit times to miss the second epoch of observations.  Additional timing data, with precision comparable to the most precise of the given data, $\sim 30s$, and at an epoch that is not commensurate with the existing gap in coverage will allow for a more complete investigation of the system.

\subsection{Constraints on Secondary Planets}

That the results of the planet search were inconclusive is not particularly surprising since the transit timings show little variation from a constant period and the point that deviates the most is suspect.  If a satisfactory solution cannot be determined from these data, we can still use them to place constraints on the orbital elements of a two-planet system.  Of particular interest is a constraint on the mass that the secondary planet could have as a function of various orbital elements; similar to the constraint that resulted from analyzing the OGLE-1998-BUL-14 microlensing event~\citep{albrow00}.  As the mass of the hypothetical perturbing planet increases, the \cs\ of the model should grow significantly regardless of the values of the remaining orbital elements.  Therefore, we systematically studied a grid of values on the semi-major axis/eccentricity plane of the perturbing planet--assuming that the orbit of TrES-1b is initially circular.  At each point we identified the mass that a perturber needs in order to produce a large deviation from the data.

At each location in the $a$/$e$ plane we set the mass of the companion to be very small ($\ll M_{\oplus}$) and, for a random value for the longitude of pericenter and the time of pericenter passage, we calculated the \cs\ of the timing residuals.  We increased the mass of the secondary planet until the \cs\ grew by some fiducial amount.  At that point we minimized the \cs\ over the longitude of pericentre and the time of pericentre passage of the secondary planet and the initial longitude of TrES-1b.  Following the minimization the \cs\ typically fell below the fiducial amount and the mass of the perturber was again increased.  However, if after the minimization the \cs\ remained above the threshold, the orbital elements of the system were recorded, the \cs\ threshold increased, and the process repeated until either a maximum mass of the perturbing planet or a maximum \cs\ was achieved.  This procedure gives the minimum \cs\ as a function of perturbing planet mass for each point in the $a$/$e$ plane.

\begin{figure}
\includegraphics[width=\fsize]{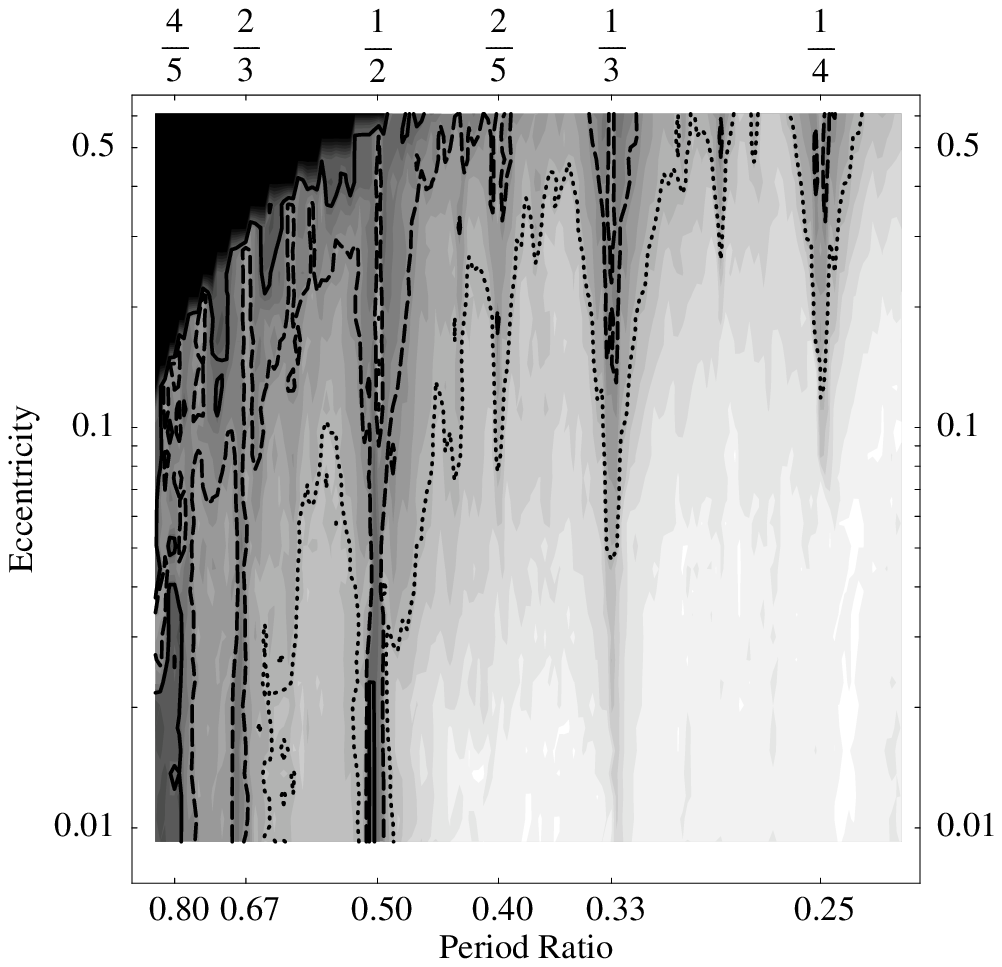}
\includegraphics[width=\fsize]{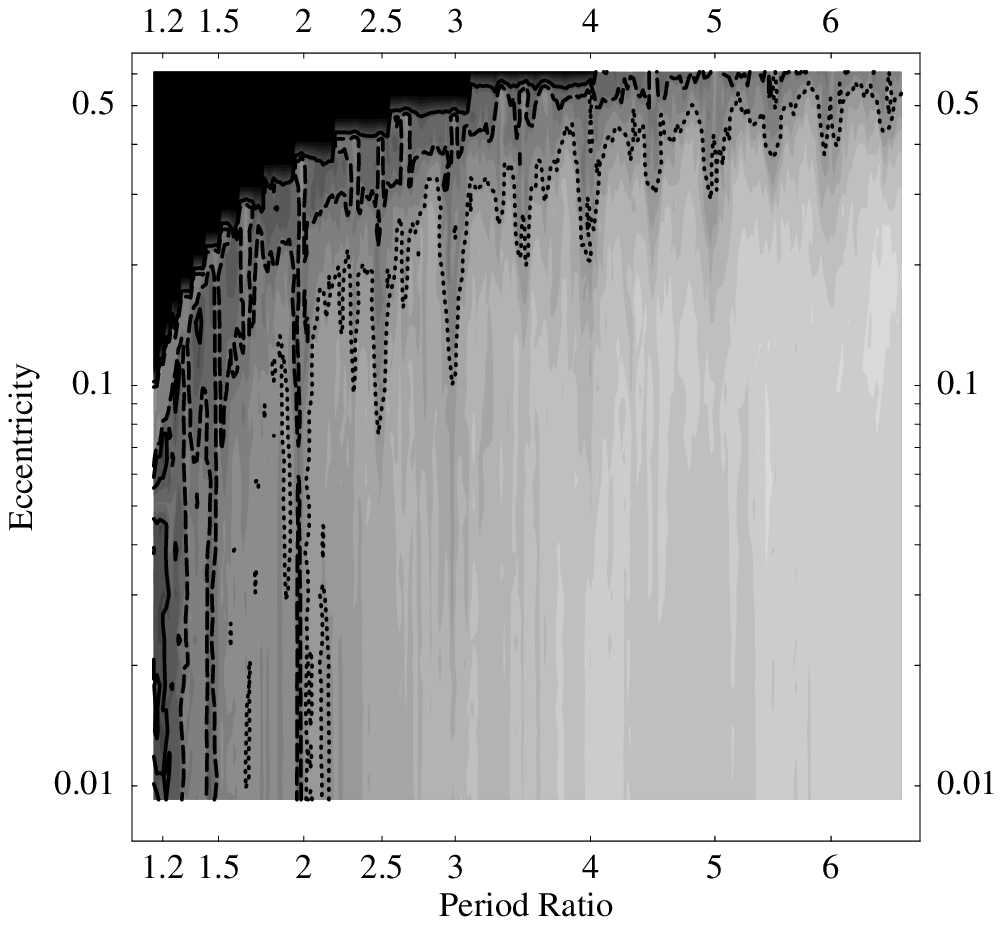}
\caption{The upper limit on the mass of an interior perturbing planet (upper panel) and exterior perturbing planet (lower panel) in the TrES-1 system from the 11-point analysis.  The regions bounded by the contours correspond to a companion mass that is less than 100 (dotted), 10 (dashed), and 1 (solid) Earth masses.}
\label{3sigin}
\end{figure}

The mass that corresponds to a 3-$\sigma$ increase in the \cs\ constitutes our estimate for the maximum allowed mass of a secondary planet.  We use the approach outlined by \cite{press} where we locate the mass that causes the difference between the \cs\ of the null hypothesis and the \cs\ obtained with a secondary planet to equal $\Delta \chi^2 = 9$.  This gives the 3-$\sigma$ limit because we allow only one parameter to vary--the remaining parameters are either fixed or marginalized at each point.  This maximum mass is shown as a function of $a$ and $e$ for the 11-point set in Figure \ref{3sigin} for both an interior and an exterior perturber.  The contours correspond to 100, 10, and 1 $M_{\oplus}$.  The dark portion in the upper-left corners are where the orbits overlap and we assign a mass of $10^{-8} M_{\odot}$ to those locations.  We see from these figures many regions where the mass of a companion must be comparable to or less than the mass of the Earth regardless of its orbital eccentricity.  The most stringent constraints are near the $j$:$j+1$ mean-motion resonances.  Of particular interest are the very tight constraints for low-eccentricity perturbers (which constraints continue to an eccentricity of zero) because tidal circularization drives the eccentricity down.  The 12-point analysis gave similar results which are not shown.

\subsection{Comparison with Radial Velocity}

\begin{figure}
\includegraphics[width=\fsize]{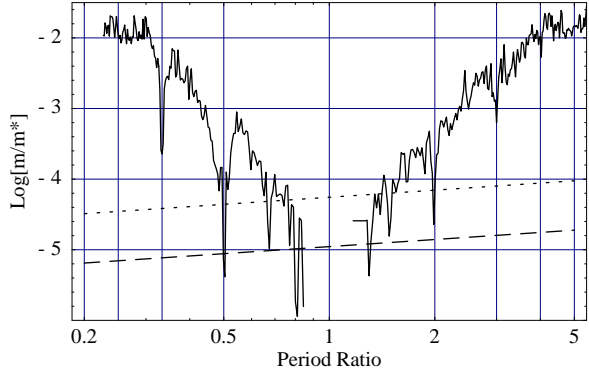}
\caption{A comparison of the limits expressed in this paper for the 11-point analysis (solid curve) and the expected limits that are achieved with 11 randomly selected radial velocity measurements that have a precision of 5 m ~s$^{-1}$ (dotted curve) and 1 m ~s$^{-1}$ (dashed curve) for a secondary planet with an eccentricity of 0.05.  The radial velocity curves are given by equation (\ref{rvform}) using $\Delta \chi^2 = 9$, $N = 11$, $Q_0 = 0$, and $Q = 2$.  Near the 4:3, 3:4, and the 1:2 mean-motion resonances the TTV analysis of these data can place tighter constraints (about a factor of 3) on the mass of putative secondary planets than can the RV technique with 1 m ~s$^{-1}$ precision.}
\label{rvcomp}
\end{figure}

\begin{figure}
\includegraphics[width=\fsize]{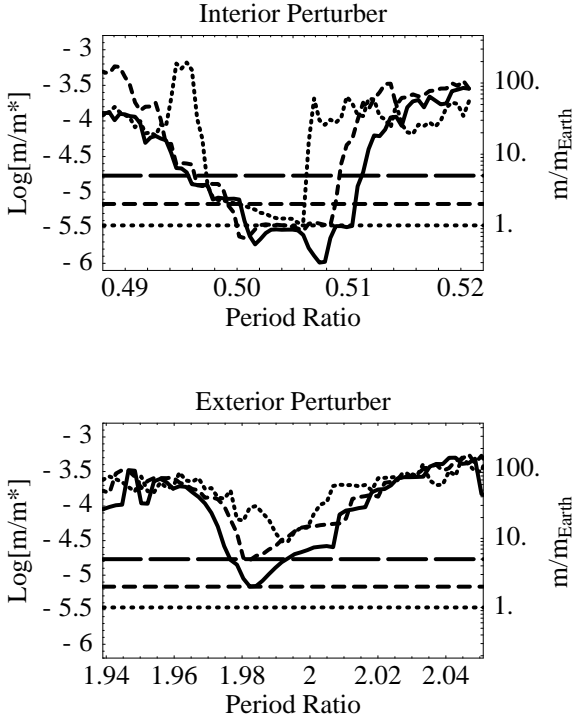}
\caption{This shows the perturber mass limit from a more refined TTV analysis near the 2:1 mean-motion resonances as a function of the period ratio of the perturber and TrES-1b.  The upper panel is for an interior perturber and the lower panel is for an exterior perturber.  The solid curve is for a perturber eccentricity of 0.00, the dashed curve is for an eccentricity of 0.02, and the dotted curve is for an eccentricity of 0.05.  The long-dashed line shows a mass of $5M_{\oplus}$, the dashed line is for $2M_{\oplus}$, and the dotted line is for $1M_{\oplus}$.  For a perturber with eccentricity 0.02, the lowest point on the curve corresponds to a mass of $0.65M_{\oplus}$.}
\label{zoom1}
\end{figure}

\begin{figure}
\includegraphics[width=\fsize]{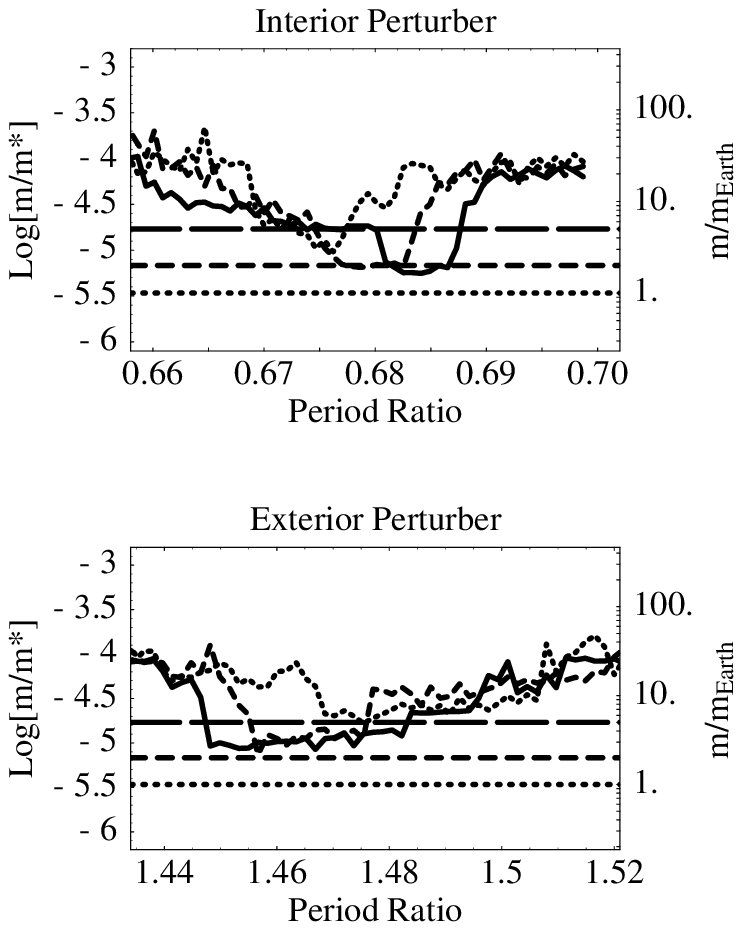}
\caption{This shows the perturber mass limit from a more refined TTV analysis near the 3:2 mean-motion resonances as a function of the period ratio of the perturber and TrES-1b.  The legend for this plot is the same as in Figure \ref{zoom1}.}
\label{zoom2}
\end{figure}

Much of the region where a secondary planet is not tightly constrained by our analysis could be limited more efficiently by radial velocity measurements.  Figure \ref{rvcomp} shows the constraint achieved from this TTV analysis and the constraint from a hypothetical RV analysis, with the same number of data, as a function of the period ratio of the two planets.  We assume that the RV measurements have a precision of both 5 and 1 m ~s$^{-1}$ and that the orbit of the perturber can be treated as circular.  Figures \ref{zoom1} and \ref{zoom2} are the same comparison focussed on the region surrounding the 2:1 and 3:2 mean-motion resonances respectively.

We derived the RV limit in these figures by assuming that the velocity residuals that remain after removing the effects of TrES-1b randomly sample the phase of the putative second planet and that the noise is uncorrelated with the RV measurements.  If no secondary planet exists, then the residuals should surround zero with a variance equal to $\sigma^2$ where $\sigma$ is the precision of the measurements.  The expected \cs\ is then equal to $\chi^2_0 = N-Q_0$ where $Q_0$ is the number of model parameters.  On the other hand, if a secondary planet exists, the expected \cs\ is
\begin{equation}
\chi^2_{\rm p} = \frac{(N-Q)}{\sigma^2} \left[ \frac{\mu^2}{2} \left( \frac{2\pi G M_0}{P} \right)^{2/3} + \sigma^2 \right]
\end{equation}
where $\mu$ is the planet to star mass ratio, $M_0$ is the mass of the star, $G$ is Newton's constant, and $Q$ is the number of model parameters.  By subtracting $\chi^2_0$ from $\chi^2_{\rm p}$ and solving for the mass ratio we obtain the mass of a secondary planet that is detectable with a given $\Delta \chi^2$ threshold as a function of the period of the planet and the precision of the radial velocity measurements
\begin{equation}\label{rvform}
\mu = \sigma \left[ \frac{2(\Delta \chi^2 + Q - Q_0)}{N-Q} \right]^{1/2} \left( \frac{P}{2\pi G M_0} \right)^{1/3}.
\end{equation}

We see from Figure \ref{rvcomp} that for low-order, mean-motion resonances the TTV analysis is able to place constraints on the mass that are nearly an order-of-magnitude smaller than the RV technique with 5 m ~s$^{-1}$ precision and that there are regions where it is more sensitive than RV measurements with 1 m ~s$^{-1}$ precision.  Meanwhile, in nonresonant regimes the RV method remains superior for a large portion of the parameter space.  Additional transit timing data, particularly with precision that is comparable to the most precise of the given data, would lower the entire limit from TTV, provided that no secondary planet exists.  Such data would render the TTV approach superior over a larger range of periods.

\section{Discussion}

Ultimately a combined analysis of all available data, including studies of the stability of candidate systems, will provide the most robust and sensitive determination of the parameters of each planetary system.  None of the planetary systems that compose the limits in Figures \ref{zoom1} and \ref{zoom2} below $10M_{\oplus}$ are stable for more than $10^6$ orbits; though stable orbits with comparable masses, periods, and eccentricities exist.  An overall stability analysis to accompany these TTV analyses was prohibitively expensive.  We estimate that for Figure \ref{3sigin} there were $\sim 10^7$ potential systems that were analysed and $\sim 10^5$ systems were analysed for each curve in Figures \ref{zoom1} and \ref{zoom2}--requiring a total of $\sim 10^8$ hours of processor time.  General stability analyses are often better suited to confirm or study a particular candidate system (e.g. \cite{rory,ji05}) than to constrain candidate systems in the manner that we have done in this section.  We did conduct a stability analysis for the system shown in Figure \ref{timings1} and found it to be unstable within $10^6$ orbits; no neighboring stable systems of comparable \cs\ were found.

It is unclear what fraction of \textit{probable} planetary companions are excluded by our results.  Aside from the fact that many of the known, multi-planetary systems are in mean-motion orbital resonance (e.g. GJ 876), recent works by \cite{thommes}, \cite{papal}, and \cite{zhou} show that, under fairly general initial conditions, small planets are readily trapped in low-order, mean-motion resonances with a gas-giant as it interacts with a protoplanetary disk and migrates inward.  These results may imply that resonant systems are common among multiple-planet systems.  If this is true, then results like Figure \ref{3sigin} actually exclude a much larger fraction of the probable orbits than one might infer from the projection of the excluded regions onto the $a$/$e$ plane.  In addition, \cite{mande} show that a large fraction of existing, terrestrial planets can survive the migration of a Jupiter-mass planet, though only a fraction of the survivors will be in resonance.

The sensitivity of TTV to the mass of a perturbing planet renders it ideal for discovering and constraining the presence of additional planets in transiting systems like TrES-1.  These studies can help determine the ubiquity of multiple planet systems and resonant systems--including the distribution of mass in those systems.  Moreover, TTV analyses of several systems can play a role in identifying the importance of various planet-formation mechanisms.  For example, the presence of close-in terrestrial planets favors a sequential-accretion model of planet formation over a gravitational instability model~\citep{zhou}.

For the case of TrES-1, while a companion planet with a mass larger than the Earth is ruled out in a portion of the available parameter space, there remain large regions where additional planets could reside; this includes mutually inclined orbits which were not studied in this work (we believe that these results are valid provided that the angle of mutual inclination $i$ satisfies $\cos i \simeq 1$).  The 11 timings analysed in this work, excepting the questionable ``12$^{\rm th}$'' point, deviate from a constant period by an amount that is difficult to interpret; having a \cs\ of about 2 per degree of freedom.  Additional observations, with higher precision, will allow us to more fully constrain the system and to interpret the existing transit timing variations.

\section*{Acknowledgments}

This work was funded in part by a Royalty Research Grant from the University of Washington.

\end{document}